\documentstyle[11pt]{article}
\begin{document}
\thispagestyle{empty}
\begin{center}
\vspace{1.8cm} {\bf REPRESENTATIONS OF GENERALIZED A$_r$
STATISTICS AND EIGENSTATES OF JACOBSON GENERATORS }\\
\vspace{1.5cm} {\bf M. Daoud}\footnote{m$_{-}$daoud@hotmail.com}\\
\vspace{0.5cm} {\it Facult\'e des Sciences, D\'epartement de
Physique, LPMC,\\
Agadir, Morocco}\\[1em]
\vspace{3cm} {\bf Abstract}
\end{center}
\baselineskip=18pt
\medskip
We investigate a generalization of $A_r$ statistics discussed
recently in the literature. The explicit complete set of state
vectors for the $A_r$ statistics system is given. We consider a
Bargmann or an analytic function description of the Fock space
corresponding to $A_r$ statistics of bosonic kind. This brings, in
a natural way, the so-called Gazeau-Klauder coherent states
defined as eigenstates of the Jacobson annihilation operators. The
minimization of Robertson uncertainty relation is also considered.

\newpage
\section{Introduction}
The generalized quantum statistics have widely discussed in the
literature (see the pioneer works [1-6]). The continuous interest
in statistics, different from Bose and Fermi statistics, is mainly
motivated by their possible relevance in the theories of
fractional quantum Hall effect [7-8], the anyon superconductivity
[9] and in high energy physics to give an adequate description of
the black hole statistics [10]. As examples of the various quantum
statistics investigated in the literature, we may cite
parastatistics introducd by Green [1-3], anyonic statistics [4],
quonic statistics [11], $k$-fermionic statistics [12] and Haldane
fractional statistics [13].\\

In the Green generalization of Bose and Fermi statistics [1], the
classical bilinear commutators or anti- commutators related to the
standard statistics are replaced by certain triple relations
satisfied by creation and annihilation operators. Now, it is well
established that the triple relations defining the parafermions,
given by
\begin{equation}
\big[[ f_i^+ , f_j^-]  , f_k^- ]\big] = - 2 \delta_{ik}
f_j^- ,{\hskip 0.5cm} \big[[ f_i^+ , f_j^+]  , f_k^- ]\big] = - 2
\delta_{ik}
f_j^+ +  2 \delta_{jk}
f_i^+, {\hskip 0.5cm} \big[[ f_i^- , f_j^-]  , f_k^- ]\big] =
0,
\end{equation}
are associated with the orthogonal Lie algebra $so(2r+1) = B_r$
$(i,j,k = 1,2,...,r)$ [14] and the trilinear commutation relations
defining the parabosons,
\begin{equation}
\big[\{ b_i^+ , b_j^-\}   , b_k^- ]\big] = - 2 \delta_{ik}
b_j^- ,{\hskip 0.5cm}\big[\{ b_i^+ , b_j^+\}   , b_k^- ]\big] = - 2
\delta_{ik}
b_j^+ - 2 \delta_{jk}
b_i^+,  {\hskip 0.5cm} \big[\{ b_i^- , b_j^-\}  , b_k^- ]\big] =
0,
\end{equation}
  are connected to the orthosymplectic superalgebra $osp(1/2r) =
B(0,r)$ [15]. Very recently, in the spirit of Palev works [6 ,
16], a classification of generalized quantum statistics were
derived for the classical Lie algebras $A_r$, $B_r$, $C_r$ and
$D_r$ [17]. In this letter, we are mainly interested by  the
generalization of the so-called $A_r$ statistics introduced by
Palev [6]. This generalization incorporate two kinds of
statistics. The first one contain statistics satisfying a
generalized exclusion Pauli principle and coincides with ones
derived by Palev [6-16]. This class will be called here fermionic
$A_r$ statistics. The second class is of bosonic kind. The
particles can be accommodated in a given quantum state without any
restriction. In the following section, we shall give the
definition of the generalized $A_r$ by means of the so-called
Jacobson generators [18]. We use the trilinear relations defining
the generalized quantum $A_r$ statistics to construct the
corresponding Fock space. We give the actions of the corresponding
creation and annihilation operators. The spectrum of the
hamiltonian, describing free particles obeying the generalized
$A_r$ statistics, is determined. In section 3, we develop  an
analytical representation corresponding to Fock space of bosonic
$A_r$ statistics. We give the differential realization of Jacobson
operators. Finally, we show that there emerge in a natural way the
so-called Gazeau-Klauder coherent states (eigenstates of Jacobson
generators) satisfying a completeness relation with respect to a
measure which will be computed explicitly. In the last section, we
show that the Gazeau-Klauder coherent states derived, in the
previous section, minimize the Robertson uncertainty relation. The
last section comprises concluding remarks.

\section{The generalized $A_r$ statistics}
In this section, we define the generalized $A_r$ statistics
through a set of $2r$ Jacobson generators satisfying certain
triple relations. We construct the Fock states and we determine
the spectrum of the Hamiltonian associated with a quantum system
obeying the generated $A_r$ statistics.\\\\ To begin, let us
consider a set of $2r$ Jacobson generators: $r$ raising operators
$x_i^+$ and $r$ lowering operators $x_i^- = (x_i^+)^{\dagger} $,
($i = 1, 2,..., r$) by means of which we define a triple Lie
system. Inspired by the para-Fermi statistics, these $2r$
operators should satisfy certain conditions and relations. First,
the creations operators are mutually commuting ($[ x_i^+ , x_j^+ ]
= 0$). A similar statement holds for the annihilation operators
($[ x_i^- , x_j^- ] = 0$). Moreover, the Jacobson generators
satisfy the following triple relations
\begin{equation}
\big[[ x_i^+ , x_j^- ] , x_k^+ ]\big] = - \epsilon  \delta_{jk}
x_i^+ - \epsilon  \delta_{ij} x_k^+
\end{equation}
\begin{equation}
\big[[ x_i^+ , x_j^- ] , x_k^- ]\big] =  \epsilon  \delta_{ik}
x_j^- + \epsilon  \delta_{ij}  x_k^-
\end{equation}
where $\epsilon \in {\bf R} $. It is interesting to note that the
Jacobson generators were initially introduced to provide an
alternative description to the Chevally one of classical Lie
algebras. This description via generators and triple relations
constitute a particular case of the definition of Lie algebras via
Lie triple systems. To be more explicit, the algebra ${\cal A}$
defined by means  of the Jacobson generators $\{x_i^+ , x_i^- , i
= 1, 2, \cdots, r \}$ and the triple relations (3) and (4) can be
re-formulated in terms of the notion of Lie triple systems.
Indeed, recall that  a vector space with trilinear composition
$[x,y,z]$ is called Lie triple system if the following identities
are satisfied:
$$[x,x,x] = 0,$$
$$[x,y,z] + [y,z,x] + [z,x,y] = 0,$$ $$[x,y,[u,v,w]] =
[[x,y,u],v,w] + [u,[x,y,v],w] + [u,v,[x,y,w]].$$\nonumber
According to this definition, the algebra ${\cal A}$ is closed
under the ternary operation $[x,y,z] = [[x,y],z]$ and define a Lie
triple system.\\ We introduce the generalized $A_r$ quantum
statistics by means of $r$ pairs of , mutually commuting, creation
annihilation operators  $a_i^{\pm} =
\frac{x_i^{\pm}}{\sqrt{|\epsilon|}}$ defined as a simple rescaling
of the Jacobson generators introduced above. The triple relations
(3) and (4) rewrite now as
\begin{equation}
\big[[ a_i^+ , a_j^- ] , a_k^+ ]\big] = - s  \delta_{jk}
a_i^+ - s  \delta_{ij} a_k^+
\end{equation}
\begin{equation}
\big[[ a_i^+ , a_j^- ] , a_k^- ]\big] =  s  \delta_{ik}
a_j^- + s \delta_{ij}  a_k^-
\end{equation}
where $ s= \frac{\epsilon}{|\epsilon|}$ is the sign of the
parameter $\epsilon$. This redefinition is more convenient for our
investigation. The new parameter $s$ play an important in
determining the Fock representation associated with the
generalized $A_r$ statistics and one obtain different microscopic
and macroscopic statistical properties according to $s = 1$ or $s
= -1$. Remark that for $s = -1$, the relations (5) and (6)
coincides with ones defining the so-called $A_r$ statistics
introduced by Palev [6].\\

To characterize a quantum gas obeying the generalized $A_r$ statistics, we
have to specify the
Hamiltonian of the system. The operators $a_i^{\pm}$ define
creation and annihilation operators for a quantum mechanical
system, described by an Hamiltonian $H$, when the Heisenberg
equation of motion
\begin{equation}
[ H , a_i^{\pm} ] = \pm e_i a_i^{\pm}
\end{equation}
is  fulfilled. The quantities $e_i$ are the energies of the modes
$i = 1, 2, ... , r$. One can verify that if $|E \rangle$ is an
eigenstate with energy $E$, $a_i^{\pm}|E \rangle$ are eigenvectors
of $H$ with energies $E \pm e_i$. In this sense, the operators $
a_i^{\pm}$ can be interpreted as ones creating or annihilating
particles. To solve the consistency equation (7), we write the
Hamiltonian $H$ as
\begin{equation}
H = \sum_{i=1}^{r} e_i h_i + c
\end{equation}
where $c$ is a constant to be specified later (It will be defined
so that the ground state energy of $H$ is vanishing). The
Hamiltonian $H$ seems to be a simple sum of "free"
(non-interacting) Hamiltonians $h_i$. However, note that, in the
quantum system under consideration, the interactions are of
statistical nature and are encoded in the triple commutation
relations. Using the structure relations of the algebra ${\cal
A}$, the solution of the Heisenberg condition (7) is given by
\begin{equation}
h_i = \frac{s}{r+1}\bigg[ (r+1)[ a_i^- , a_i^+ ] - \sum_{j=1}^{r}[ a_j^- ,
a_j^+]
\bigg].
\end{equation}

We now consider an Hilbertean representation of the algebra ${\cal
A}$. Let  ${\cal F}$ be the Hilbert-Fock space on which the
generators of ${\cal A}$ act. Since, the algebra ${\cal A}$ is
spanned by $r$ pairs of Jacobson generators, it is natural to
assume that the Fock space is given by
\begin{equation}
{\cal F} = \{ |n_1, n_2,\cdots , n_r\rangle\ , n_i \in {\bf N} \}.
\end{equation}
We define the action of $a_i^{\pm}$ on ${\cal F}$ as
\begin{equation}
a_i^{\pm} |n_1,\cdots, n_i,\cdots , n_r\rangle\ = \sqrt{F_i
(n_1,\cdots ,n_i \pm 1,\cdots, n_r)}|n_1,\cdots, n_i \pm 1,\cdots
, n_r\rangle\
\end{equation}
in terms of the functions $F_i$ called the structure functions
which should be non-negatives so that all states are well defined.
To determine the expressions of the functions $F_i$ in terms of
the quantum numbers $n_1, n_2, \cdots, n_r$, let first assume the
existence of a vacuum vector $a_i^- |0, 0,\cdots , 0\rangle\ = 0$
for all $i = 1, 2, \cdots, r$. This implies that the functions
$F_i$ can be written
\begin{equation}
F_i (n_1,\cdots ,n_i,\cdots, n_r) = n_i \Delta F_i (n_1,\cdots
,n_i,\cdots, n_r)
\end{equation}
in a factorized form where the new functions $\Delta F_i$ are
defined such that $\Delta F_i (n_1,\cdots ,n_i = 0,\cdots,
n_r)\neq 0$ for $i=1, 2,\cdots, r$. Furthermore, since the
Jacobson operators satisfy the trilinear relations (5) and (6),
these functions should be linear in the quantum numbers $n_i$:
\begin{equation}
\Delta F _i (n_1,\cdots ,n_i,\cdots, n_r)= k_0 + (k_1 n_1 + k_2
n_2 + \cdots + k_r n_r ).
\end{equation}
Finally, using the relations $[a_i^+ , a_j^+] = 0$ and $\big[[
a_i^+ , a_i^- ] , a_i^+ ]\big] = - 2s a_i^+ $, one verify $k_i =
k_j $ and $k_i = s$, respectively. For convenience, we set  $k_0 =
k - \frac{1+s}{2}$ assumed to be a non vanishing integer. The
actions of Jacobson generators on the
states spanned the Hilbert-Fock space ${\cal F}$ are now given by
\begin{equation}
a_i^{-} |n_1,\cdots, n_i,\cdots , n_r\rangle\ = \sqrt{n_i (k_0 +
s(n_1+n_2+\cdots+n_r))}|n_1,\cdots, n_i-1,\cdots , n_r\rangle\
\end{equation}
\begin{equation}
a_i^{+} |n_1,\cdots, n_i,\cdots , n_r\rangle\ = \sqrt{(n_i+1) (k_0
+ s(n_1+n_2+\cdots+n_r+1))}|n_1,\cdots, n_i+1,\cdots , n_r\rangle.
\end{equation}
The dimension of the irreducible representation space ${\cal F}$
is determined by the  condition:
\begin{equation}
k_0 + s(n_1+n_2+\cdots+n_r) > 0.
\end{equation}
It depends on the sign of the parameter $s$. It is clear that for
$s=1$, the Fock space ${\cal F}$ is infinite dimension. However,
for $s=-1$, there exists a finite number of states satisfying the
condition $n_1+n_2+\cdots+n_r \leq k-1$. The dimension is given,
in this case, by $\frac{(k-1+r)!}{(k-1)!r!}$. This is exactly the
dimension of the Fock representation of $A_r$ statistics discussed
in [6]. This condition-restriction is closely related to so-called
generalized exclusion Pauli principle according to which no more
than $k-1$ particles can be accommodated in the same quantum
state. In this sense, for $ s = -1$, the generalized $A_r$ quantum
statistics give statistics of fermionic behaviour. They will be
termed here as fermionic $A_r$ statistics and ones corresponding
to $s = 1$ will be named bosonic $A_r$ statistics.\\
Setting $c =
\frac{r}{r+1} s k_0$ in (8) and using the equation (9) together
with the actions of creation and annihilation operators (14-15),
one has
\begin{equation}
H |n_1,\cdots, n_i,\cdots , n_r\rangle\ = \sum_{i=1}^{r}e_i
n_i|n_1,\cdots, n_i,\cdots , n_r\rangle\,
\end{equation}
the discrete spectrum of the Hamiltonian $H$ associated to $A_r$
statistics systems. Finally, we point out one interesting property
of the generalized $A_r$ statistics. Introduce the operators
$b_i^{\pm} = \frac{a_i^{\pm}}{\sqrt{k}}$ for $i = 1, 2, \cdots, r$
and consider $k$ very large. From equations (14) and (15), we
obtain
\begin{equation}
b_i^{-} |n_1,\cdots, n_i,\cdots , n_r\rangle\ =
\sqrt{n_i}|n_1,\cdots, n_i-1,\cdots , n_r\rangle\
\end{equation}
\begin{equation}
b_i^{+} |n_1,\cdots, n_i,\cdots , n_r\rangle\ =
\sqrt{n_i+1}|n_1,\cdots, n_i+1,\cdots , n_r\rangle.
\end{equation}
In this limit, the generalized $A_r$ statistics (fermionic and
bosonic ones) coincide with th Bose statistics and the Jacobson
operators reduce to Bose ones (creation and annihilation operators
of harmonic oscillators).\\ Beside the Fock representation, it is
interesting to look for an analytical realization of the space
representation associated with the generalized $A_r$ statistics.
This realization can constitute an useful tool to construct a path
integral formula to describe the quantum dynamics of the system
described by the Hamiltonian $H$. In the following section, we
treat this question in the particular case of bosonic $A_r$
statistics.

\section{Bargmann realization of  bosonic $A_r$ statistics}
In this section, we give a realization \`a la Bargmann using a
suitably defined Hilbert space of entire analytic functions
associated to the  bosonic $A_r$ statistics introduced above ($s =
1 $). In this analytic realization, the Jacobson creation
operators are realized as simple multiplication by some complex
variables. As by product, this realization brings, in a natural
way, the Gazeau-Klauder  coherent states [19] associated to a
quantum mechanical system described by the Hamiltonian given by
(8) and (9).\\ To start, we ask for a realization in which the
vector $|k;n_1,\cdots,n_r\rangle$ is realized as powers of complex
variables $\omega_1,\cdots,\omega_r$
\begin{equation}
|k; n_1,\cdots,n_r\rangle \longrightarrow C_{k; n_1,\cdots,n_r}
\omega_1^{n_1}\cdots\omega_r^{n_r},
\end{equation}
such that Jacobson creation operators $a_i^+$act as a simple
multiplication by $\omega_i$. Using the actions of the creation
operators, one verify that  the  coefficients $C_{k;
n_1,\cdots,n_r}$, occurring in (20), satisfy the following
recursion formula
\begin{equation}
C_{k; n_1,\cdots,n_i,\cdots,n_r} =
((n_i+1)(k+n_1+\cdots+n_i+\cdots+n_r))^{\frac{1}{2}}
C_{k;n_1,\cdots,n_i+1,\cdots,n_r}.
\end{equation}
Solving this relation, we obtain
\begin{equation}
C_{k; n_1,\cdots,n_i,\cdots,n_r} = \bigg[{\frac{(k-1 + n - n_i)!
}{n_i!(k-1 + n) !}}\bigg]^{\frac{1}{2}} C_{k;
n_1,\cdots,0,\cdots,n_r}
\end{equation}
where $n = n_1 + n_2 + \cdots + n_r$. We repeat this procedure for
all $i = 1, 2,\cdots,r$ and setting $C_{k; 0,\cdots,0} = 1$, we
obtain
\begin{equation}
C_{k; n_1,\cdots,n_i,\cdots,n_r} = \bigg[{\frac{(k-1)!
}{n_1!\cdots n_r!(k-1 + n) !}}\bigg]^{\frac{1}{2}}.
\end{equation}
If we define the operators $N_i$ ($\neq a_i^+ a_i^-$) such that
\begin{equation}
N_i|k;n_1,\cdots,n_i\cdots,n_r\rangle = n_i
|k;n_1,\cdots,n_i\cdots,n_r\rangle,
\end{equation}
it is easy to see that the operators $N_i$ act in this
differential realization as
\begin{equation}
N_i \longrightarrow \omega_i \frac{d}{d\omega_i}.
\end{equation}
To write the differential actions of the annihilation operators
$a_i^-$, we use their actions on the Fock space (Eq. 14) together
with the equation (25). We show that $a_i^-$
\begin{equation}
a_i^- \longrightarrow k \frac{d}{d\omega_i} + \omega_i
\frac{d^2}{d^2\omega_i} + \frac{d}{d\omega_i}\sum_{i \neq
j}\omega_j \frac{d}{d\omega_j}
\end{equation}
acts as a second order differential operator. A general vector
\begin{equation}
|\psi\rangle = \sum_{n_1,\cdots,n_r}
\psi_{n_1,\cdots,n_r}|k;n_1,\cdots,n_r\rangle
\end{equation}
in the Fock space ${\cal F}$ now is realized as follows
\begin{equation}
\psi(\omega_1,\cdots,\omega_r) = \sum_{n_1,\cdots,n_r}
\psi_{n_1,\cdots,n_r}C_{k;n_1,\cdots,n_r}\omega_1^{n_1}\cdots\omega_r^{n_r}.
\end{equation}
We define the inner product in the following manner
\begin{equation}
\langle\psi'|\psi\rangle = \int d^2\omega_1 \cdots d^2\omega_r
K(k;\omega_1,\cdots,\omega_r)\psi'^{\star}(\omega_1,\cdots,\omega_r)
\psi(\omega_1,\cdots,\omega_r)
\end{equation}
where $d^2\omega_i \equiv dRe\omega_idIm\omega_i$ and the
integration extends over the entire complex space $\bf{C}^r$. To
compute the measure function, appearing in the definition of the
inner product (29), we choose  $|\psi\rangle$(resp.
$|\psi'\rangle$) to be the vector $|k;n_1,\cdots,n_r\rangle$
(resp. $|k;n'_1,\cdots,n'_r\rangle$). We also assume that it
depends only on $\rho_i = |\omega_i|$ for $i = 1,\cdots, r$. This
assumption is nothing but the isotropic condition used usually in
the moment problems. It is a simple matter of computation to show
that the function $K(k;\rho_1,\cdots,\rho_r)$ should satisfy the
integral
equation
\begin{equation}
(2\pi)^r\int_0^{\infty}\cdots\int_0^{\infty} d\rho_1\cdots d\rho_r
K(k;\rho_1,\cdots ,\rho_r |\rho_1|^{2n_1+1}\cdots
|\rho_r|^{2n_r+1}= \frac{n_1!\cdots n_r!(k-1+n)!}{(k-1)!}.
\end{equation}
A solution of this equation exists and can be computed using the
inverse Mellin transform [20] (see also a nice proof in [21]) in
term of the Bessel function
\begin{equation}
K(k;R) = \frac{2}{\pi^r(k-1)!}R^{k-r}K_{k-r}(2R)
\end{equation}
where $R^2 = \rho_1^2 + \cdots + \rho_r^2$. Note that the analytic
function $\psi (\omega_1, \cdots ,\omega_r)$ can be viewed as the
inner product of the ket $|\psi\rangle$ with a bra $\langle k;
\omega_1^{\star}, \cdots ,\omega_r^{\star}|$ labeled by the
complex conjugate of the variables $\omega_1, \cdots ,\omega_r$
\begin{equation}
\psi (\omega_1, \cdots ,\omega_r) = {\cal N}\langle k;
\omega_1^{\star}, \cdots ,\omega_r^{\star}|\psi \rangle
\end{equation}
where ${\cal N} \equiv {\cal N}(|\omega_1|, \cdots ,|\omega_r|)$
stands for a normalization constant of the states $|k; \omega_1,
\cdots ,\omega_r\rangle$ to be specified later. For the particular
case $|\psi \rangle = |k; n_1,\cdots,n_r\rangle$, we get
\begin{equation}
\langle k; \omega_1^{\star}, \cdots ,\omega_r^{\star}|k;
n_1,\cdots,n_r \rangle = {\cal N}^{-1} C_{k;n_1,\cdots,n_r}
\omega_1^{n_1} \cdots \omega_r^{n_r}.
\end{equation}
The last equation implies
\begin{equation}
|k; \omega_1, \cdots ,\omega_r\rangle = {\cal N}^{-1}
\sum_{n_1=0}^{\infty} \cdots \sum_{n_r=0}^{\infty}
\bigg[{\frac{(k-1)! }{n_1!\cdots n_r!(k-1 + n)
!}}\bigg]^{\frac{1}{2}} \omega_1^{n_1} \cdots \omega_r^{n_r}|k;
n_1,\cdots,n_r\rangle.
\end{equation}
The normalization constant ${\cal N}$ is easily evaluated. It is
given by
\begin{equation}
{\cal N}^2(|\omega_1|, \cdots ,|\omega_r|) = \sum_{n_1=0}^{\infty}
\cdots \sum_{n_r=0}^{\infty} {\frac{(k-1)! }{n_1!\cdots n_r!(k-1 +
n) !}} |\omega_1|^{2n_1} \cdots |\omega_r|^{2n_r}.
\end{equation}
It may be noted that the states $|k; \omega_1, \cdots
,\omega_r\rangle$ are not orthogonal. They constitute an
over-complete set with respect to the measure given by (31). It is
also interesting to remark that they are eigenvectors of the
Jacobson operators $a_i^-$ with the eigenvalues $\omega_i$. Since
the Gazeau-Klauder coherent states [19], usually associated with
exactly sovable quantum mechanical systems and defined as
eigenstates of the annihilation operator, the states $|k;
\omega_1, \cdots ,\omega_r\rangle$ can be considered as ones of
Gazeau-Klauder type. They are associated with the system described
the Hamiltonian (8) and obeying the bosonic $A_r$ statistics.
\section{Robertson uncertainty relation}
The main aim of this section is to show that the Gazeau-Klauder
coherent states, derived in the previous section, minimize the
Robertson uncertainty relation. Note that the states minimizing
this relation are called intelligent states. Recall that, for $2r$
observables (hermitian operators) $(X_1,X_2, \cdots , X_{2r} )
\equiv X$, Robertson [22] (see also [23-24] and references
therein) established the following uncertainty relation for the
matrix dispersion $\sigma$
\begin{equation}
\det \sigma(X) \geq \det C(X)
\end{equation}
where $\sigma_{\alpha \beta} = \frac{1}{2}\langle
X_{\alpha}X_{\beta} + X_{\beta}X_{\alpha} \rangle - \langle
X_{\alpha}X_{\beta} \rangle$, $(\alpha = 1, 2, \cdots , 2r )$, and
$C$ is the antisymmetric matrix of the mean commutators $C_{\alpha
\beta} = -\frac{i}{2}[X_{\alpha} , X_{\beta}]$. Here $\langle O
\rangle$ stands for the mean value of the operator $O$ in a given
quantum state which is generally a mixed state. For $r=1$,
inequality (36) coincides with Shr\"odinger uncertainty relation
which gives the Heisenberg uncertainty relation when the
covariance $\sigma_{12}$ is vanishing. To show that the
Gazeau-Klauder coherent states minimize the uncertainty relation
(36),i.e. $\det \sigma(X) =  \det C(X)$, let define the hermitian
operators $(X_1,X_2, \cdots , X_{2r})$ as
\begin{equation}
X_i = \frac{1}{2} (a_i^+ + a_i^-) {\hskip 1cm} X_{i+r} =
\frac{i}{2} (a_i^+ - a_i^-)
\end{equation}
in terms of the creation and annihilation Jacobson generators.\\
The matrix $A \equiv (a_1^-, a_2^-, \cdots, a_r^-, a_1^+, a_2^+,
\cdots, a_r^+)$ is related to $X$ as $X = U A$
\begin{displaymath}
U = \frac{1}{2}\left( \begin{array}{ccc} {\bf 1}_r & {\bf 1}_r \\
-i {\bf 1}_r & i {\bf 1}_r
\\
\end{array} \right)
\end{displaymath}
where ${\bf 1}_r$ is $r\times r$ unit matrix. It follows that both
matrices  $\sigma (X)$ and $C(X)$  can be expressed in terms of
matrices  $\sigma (A)$ and $C(A)$:
\begin{equation}
\sigma (X) = U \sigma (A) U^T {\hskip 1cm} C(X) = U C(A) U^T.
\end{equation}
At this step, one can evaluate the matrix elements of $\sigma (A)$
and $C(A)$. Indeed, using the eigenvalue equations $a_i^- |k;
\omega_1, \cdots, \omega_i, \cdots ,\omega_r\rangle = \omega_i |k;
\omega_1, \cdots, \omega_i, \cdots ,\omega_r\rangle$, one has
\begin{equation}
\sigma_{ij} = 0 {\hskip 1cm} C_{ij} = 0
\end{equation}
\begin{equation}
\sigma_{i+r,j+r} = 0 {\hskip 1cm} C_{i+r,j+r} = 0
\end{equation}
\begin{equation}
\sigma_{i,j+r} =  i C_{i,j+r} {\hskip 1cm} \sigma_{i+r,j} = -i
C_{i+r,j}.
\end{equation}
From the last relations among the matrix elements of dispersion
and covariance matrices , we show $\det \sigma(A) = \det C(A)$.
Moreover, in view of equations (38) (and the nondegeneracy of $U$,
$\det U = {(\frac{i}{2})}^r $ ) we have $\det \sigma(X) = \det
C(X)$. Finally, we conclude that the Gazeau-klauder coherent
states, associated with a quantum system obeying the bosonic $A_r$
statistics, minimize the Robertson uncertainty relation.
\section{Concluding remarks}
To conclude, let summarize the main points discussed in this
letter. We have studied the Fock representations of the
generalized $A_r$ statistics. It contain bosonic and femionic
statistics. It is remarkable that in the limit $k$ very large ($k$
index labeling the irreducible Fock representations), bosonic as
well as fermionic statistics reduce to Bose statistics. In this
limit, the Jacobson generators coincide with the creation and
annihilation of conventional bosonic degrees of freedom. We also
obtained a Bargmann analytic realization associated to bosonic
$A_r$ statistics. An over-complete set of states is constructed.
This is excatly the set of Gazeau-Klauder coherent states of a
quantum system described by the Hamiltonian $H$ (cf. equation
(8)). Finally, we have shown that these states saturate the
Robertson uncertainty relation. Although we have performed the
analytic realization for bosonic $A_r$ statistics, similar
treatment can be done for fermionic case. Also, as continuation,
it would be useful to investigate, utilizing tools given here, the
properties of quantum statistics associated with other classical
Lie algebras and super-algebras.

\end{document}